\documentclass[sn-basic]{sn-jnl}

\usepackage{graphicx}%
\usepackage{multirow}%
\usepackage{amsmath,amssymb,amsfonts}%
\usepackage{amsthm}%
\usepackage{mathrsfs}%
\usepackage[title]{appendix}%
\usepackage{xcolor}%
\usepackage{textcomp}%
\usepackage{manyfoot}%
\usepackage{booktabs}%
\usepackage{algorithm}%
\usepackage{algorithmicx}%
\usepackage{algpseudocode}%
\usepackage{listings}%

\newcommand{\Add}[1]{\textcolor{black}{#1}}

\theoremstyle{thmstyleone}%
\theoremstyle{thmstyletwo}%

\theoremstyle{thmstylethree}%

\begin{document}

\title[Coronal Mass Ejections from Young Suns: An Observational View through the Solar--Stellar Connection]
{Coronal Mass Ejections from Young Suns: An Observational View through the Solar--Stellar Connection}

\author*[1,2,3,4]{\fnm{Kosuke} \sur{Namekata}}\email{namekata@kusastro.kyoto-u.ac.jp}

\affil*[1]{\orgdiv{The Hakubi Center for Advanced Research}, \orgname{Kyoto University}, \orgaddress{\street{Yoshida-Honmachi, Sakyo-ku}, \city{Kyoto}, \postcode{606-8501}, \state{Kyoto}, \country{Japan}}}

\affil[2]{\orgdiv{Department of Physics}, \orgname{Kyoto University}, \orgaddress{\street{Kitashirakawa-Oiwake-cho, Sakyo-ku}, \city{Kyoto}, \postcode{606-8502}, \state{Kyoto}, \country{Japan}}}

\affil[3]{\orgdiv{Heliophysics Science Division}, \orgname{NASA Goddard Space Flight Center}, \orgaddress{\street{8800 Greenbelt Road}, \city{Greenbelt}, \postcode{20771}, \state{MD}, \country{USA}}}

\affil[4]{\orgdiv{Department of Physics}, \orgname{The Catholic University of America}, \orgaddress{\street{620 Michigan Avenue, N.E.}, \city{Washington}, \postcode{20064}, \state{DC}, \country{USA}}}


\abstract{
Kepler and TESS have revealed that magnetically active stars frequently produce flares and superflares, raising the possibility that frequent, fast, and massive coronal mass ejections (CMEs) may affect planetary space weather environments. 
However, observational constraints on stellar CMEs remain extremely limited. 
Several diagnostics, including Doppler-shifted spectral lines, coronal dimming, and radio bursts, have been explored mainly for cool dwarfs, while recent observations have begun to extend these searches to young solar-type stars, which serve as important benchmarks for the solar--stellar connection. 
The goal of this review is to summarize the current understanding and recent observational progress on CME signatures from young solar-type stars, with a primary focus on Doppler-shift methods. 
It examines how these signatures resemble solar observations and models, what the inferred physical properties can tell us about the occurrence of stellar CMEs and their contribution to stellar mass loss, and how Doppler-shift signatures can be connected with multi-wavelength diagnostics. 
Comparisons with CME candidates from cool dwarfs and close binaries further place these signatures in the broader context of stellar CMEs in different stellar environments. 
These results have important implications for the space weather environments of young planets, while also pointing to key directions for future observations and modeling.
}




\maketitle

\section{Introduction}

Solar and stellar flares are sudden releases of magnetic energy in the atmosphere, observed as intense brightenings from radio to X-ray.
They are generally understood as a consequence of magnetic reconnection, where free magnetic energy stored in the active region coronae is converted into thermal, nonthermal, kinetic, and radiation energy \citep{2011LRSP....8....6S,2017LRSP...14....2B,2024LRSP...21....1K}. 
In the case of the Sun, large solar flares are often accompanied by eruptions of filaments or prominences and by coronal mass ejections (CMEs), carrying magnetized plasma from the solar corona into interplanetary space \citep{2003ApJ...586..562G,2021LRSP...18....4T}. 
These CMEs can drive shocks and energetic particles, and are therefore one of the most important sources of solar space weather \citep{2022LRSP...19....2C,2023LRSP...20....2U}.

A natural question is whether similar eruptions occur on other stars. 
With the discovery of thousands of exoplanets by Kepler and TESS, this question has become directly relevant to extra-solar space weather environments.
Solar-type stars\footnote{In this review, solar-type stars are broadly defined as G-type main-sequence stars.} are particularly important because they allow us to examine the environments of Earth-like habitable planets and to place the Sun in the broader context of the \textit{Sun-in-time} concept \citep{2007LRSP....4....3G}. 
In particular, young solar-type stars provide empirical clues to the early Solar System, when planetary atmospheres were evolving and conditions related to the emergence of life on Earth were being established. 
If these stars produce CMEs in association with large flares, young planets would be exposed not only to enhanced X-ray and ultraviolet (UV) radiation, but also to dense magnetized plasma, shocks, and energetic particles. 
These effects may influence atmospheric escape, atmospheric chemistry, and long-term planetary evolution \citep{2016NatGe...9..452A,2020IJAsB..19..136A,2025kiss.rept.....L} (see Section~\ref{sec:implication}).

\begin{figure}
    \centering
    \includegraphics[width=1.00\linewidth]{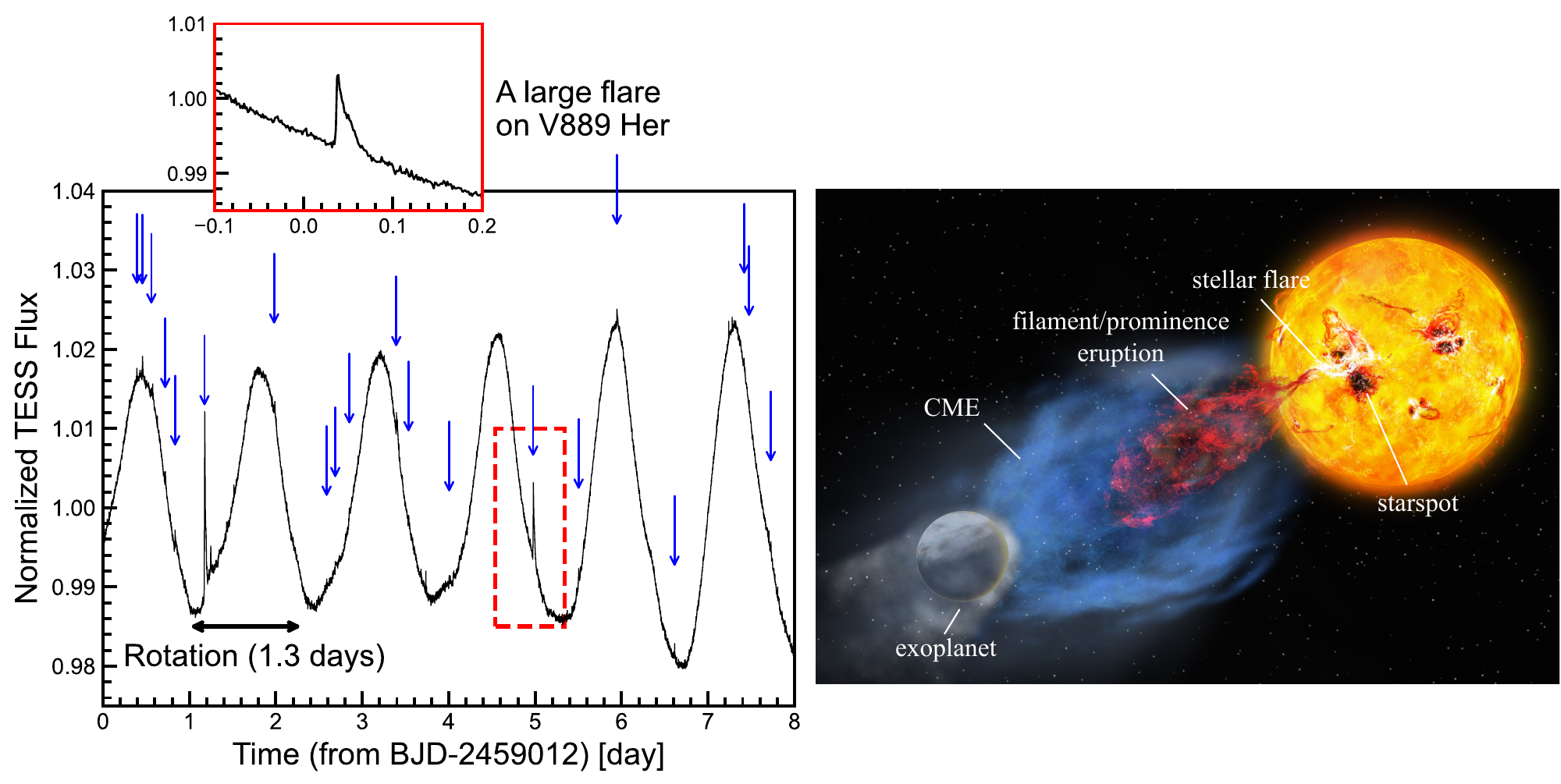}
\caption{
Left: Example of superflares on the young solar-type star V889 Her observed by TESS. Data from \citet{2025ApJ...993...80N}. Blue arrows are detected superflares.
Right: Illustration of the possible influence of a superflare on an orbiting exoplanet. Credit: NAOJ.
}
    \label{fig:flare}
\end{figure}

The importance of this problem has increased with the discovery of stellar superflares. Kepler, TESS, and even previous X-ray observatories have shown that solar-type stars can produce flares with energies exceeding $10^{33}$ erg, much larger than the largest solar flares on record ($\sim10^{33}$ erg; \citealt{2012ApJ...759...71E,2022LRSP...19....2C}), and that such events occur more frequently on young and rapidly rotating stars \citep{2007LRSP....4....3G,2012Natur.485..478M,2019ApJ...876...58N,2020arXiv201102117O,2020ApJ...890...46T,2022ApJ...926L...5N,2022PASJ...74.1295Y,2024Sci...386.1301V}. 
Figure \ref{fig:flare} shows an example of superflares on the young solar-type star V889 Hercules.
On the basis of solar flare-CME relations, it is expected that these superflares may be accompanied by frequent, fast and massive CMEs on solar-like atmospheres \citep{2009IAUS..257..233Y,2012ApJ...760....9A,2013ApJ...764..170D,2015ApJ...809...79O}. However, it has been suggested that this extrapolation may not be straightforward. Active stars have stronger and more complex magnetic fields than the present-day Sun, and such fields may either produce more energetic eruptions or suppress them through magnetic confinement \citep{2018ApJ...862...93A,2022MNRAS.509.5075S,2024MNRAS.533.1156S}. Therefore, the occurrence of a superflare does not by itself guarantee the occurrence of an escaping CME.

Observationally, stellar CMEs are still difficult to establish (see Section \ref{sec:observational_diagnostics} and many reviews by \citealt{2017IAUS..328..243O,2022SerAJ.205....1L,2022arXiv221105506N,2024Univ...10..313V,2025LRSP...22....2V,2025kiss.rept.....L,2026IAUS..388..154N}). Unlike solar CMEs, they cannot be spatially resolved or directly imaged with current techniques. Their signatures must instead be inferred from disk-integrated spectra and light curves, such as Doppler-shifted chromospheric lines, coronal dimming, and radio bursts. Each diagnostic has its own limitation, and a single signature is rarely sufficient to determine whether a CME has escaped from the stellar corona.

This article reviews recent observational progress in the search for stellar CMEs associated with superflares, with a particular focus on young solar-type stars. 
\Add{A complimentary, broader numerical perspective on stellar CMEs and exoplanets is reviewed by \citet{2022AN....34310100A}.}
Section~\ref{sec:observational_diagnostics} summarizes observational diagnostics of unresolved stellar CMEs. 
Section~\ref{sec:discovery_filament_prominence} discusses recent detections of filament and prominence eruptions from EK Draconis, while Sections~\ref{sec:sun_as_a_star} and \ref{sec:velocity_mass_kinetic_energy} describe their interpretation through the solar--stellar connection and physical property estimates. 
Sections~\ref{sec:data_driven_modeling}--\ref{sec:multi_wavelength} review data-driven modeling, occurrence rates, magnetic environments, and multi-wavelength signatures. 
Sections~\ref{sec:implication} and \ref{sec:future} discuss implications for young planetary environments and future directions in stellar CME studies.

\section{Observational diagnostics of stellar coronal mass ejections}\label{sec:observational_diagnostics}

\begin{figure}
    \centering
    \includegraphics[width=0.9\linewidth]{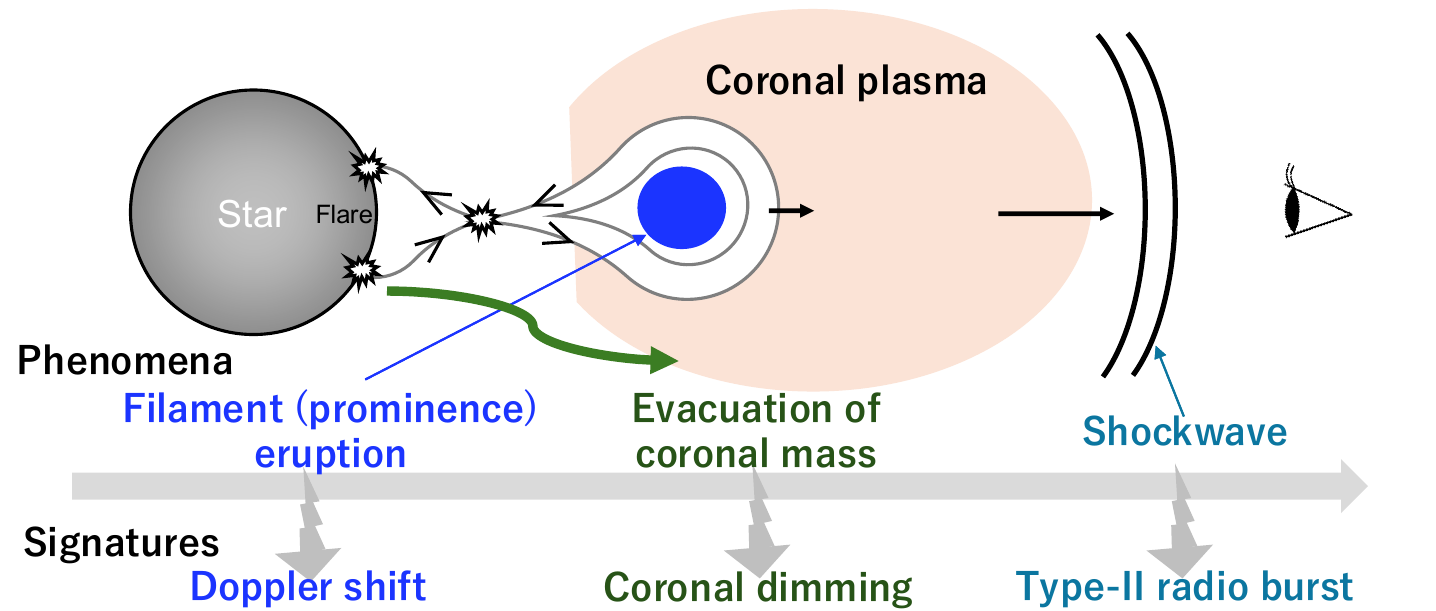}
\caption{
Schematic explanations of representative detection methods for stellar CMEs. 
}
    \label{fig:overall}
\end{figure}

A variety of indirect observational signatures have been developed and tested. A comprehensive review is given by \citet{2025kiss.rept.....L}. Figure~\ref{fig:overall} summarizes some of the most commonly used methods for detecting stellar CMEs, motivated by solar observations.

Among these methods, one of the most practical and successful approaches is to search for transient Doppler-shifted spectral lines during stellar flares. Chromospheric lines, especially H$\alpha$, are useful because cool filament or prominence material can produce blue-shifted absorption or emission components \citep[e.g.,][]{2021EP&S...73...58S,2022ApJ...939...98O}\footnote{Following solar terminology, cool material seen against the stellar disk is referred to as a filament, while material seen outside the stellar disk, without a background source, is referred to as a prominence. These are essentially the same physical structure, and the difference is mainly observational geometry.}. 
On the Sun, erupting filaments and prominences often constitute the cool core of CMEs. They are typically observed during the early phase of CME development, while the CME front can propagate several times faster \Add{\citep[e.g.,][]{2003ApJ...586..562G,2017ApJ...834..172J}}. Filament eruptions can also drain back along magnetic loops and do not always lead to escaping CMEs. Nevertheless, fast and large-scale filament/prominence eruptions are more likely to be associated with CMEs \citep[e.g.,][]{2021EP&S...73...58S}.

The Doppler method has been applied to a wide range of active stars. Blue-shifted optical emission components have been reported mostly for M dwarfs \citep[e.g.,][]{1990A&A...238..249H,2016A&A...590A..11V,2020PASJ..tmp..253M,2019A&A...623A..49V,2023ApJ...948....9I,2024PASJ...76..175I,2024ApJ...961..189N,2025ApJ...979...93K}, but their interpretation remains debated because their atmospheric conditions differ from those of the Sun. In many cases, these signatures appear only as blue-shifted emission, not absorption \citep[probably due to the faint background, see][]{2022MNRAS.513.6058L}, and their velocities are relatively low \citep{2020PASJ..tmp..253M,2024ApJ...961..189N}. 
They are sometimes interpreted as signatures of prominence eruptions. However, alternative explanations have also been discussed, including cool upward flows near flare footpoints associated with chromospheric evaporation \citep{2018PASJ...70..100T,2018PASJ...70...62H}, or red-wing absorption from post-flare loops \citep{2018PASJ...70...62H}.

Other diagnostics probe different parts of an eruption, as summarized in Figure~\ref{fig:overall}. For example, coronal dimming in X-rays or UV can indicate the evacuation of coronal plasma and is strongly associated with CMEs on the Sun \citep{2016SoPh..291.1761H}. 
In recent years, possible dimming events have been reported for M and K dwarfs and are regarded as promising CME candidates \citep{2021NatAs...5..697V,2022ApJ...936..170L,2025LRSP...22....2V}. 
In addition, in the case of the Sun, type-II and type-IV radio bursts trace shockwave propagation and energetic electrons in erupting flux ropes related to CMEs, respectively. Although stellar detections remain rare and difficult to interpret, recent efforts have reported possible evidence for type-IV \citep{2020ApJ...905...23Z,2024A&A...686A..51M} and type-II bursts \citep{2025Natur.647..603C,2025A&A...703A.198K} from M dwarfs.

In this way, several complementary methods have been explored, and promising evidence for stellar CMEs has begun to emerge, mainly for M dwarfs. In contrast, observational constraints for solar-type stars remain extremely limited. In addition, because each method is incomplete, robust identification of stellar CMEs requires comparison among multiple diagnostics and careful calibration with solar observations.
The following sections review recent progress and current status in stellar CME studies, focusing mainly on Doppler-shift methods that have provided the starting point for observational studies of CMEs from young solar-type stars.

\section{Discovery of stellar filament/prominence eruptions}
\label{sec:discovery_filament_prominence}

\begin{figure}
    \centering
    \includegraphics[width=1.0\linewidth]{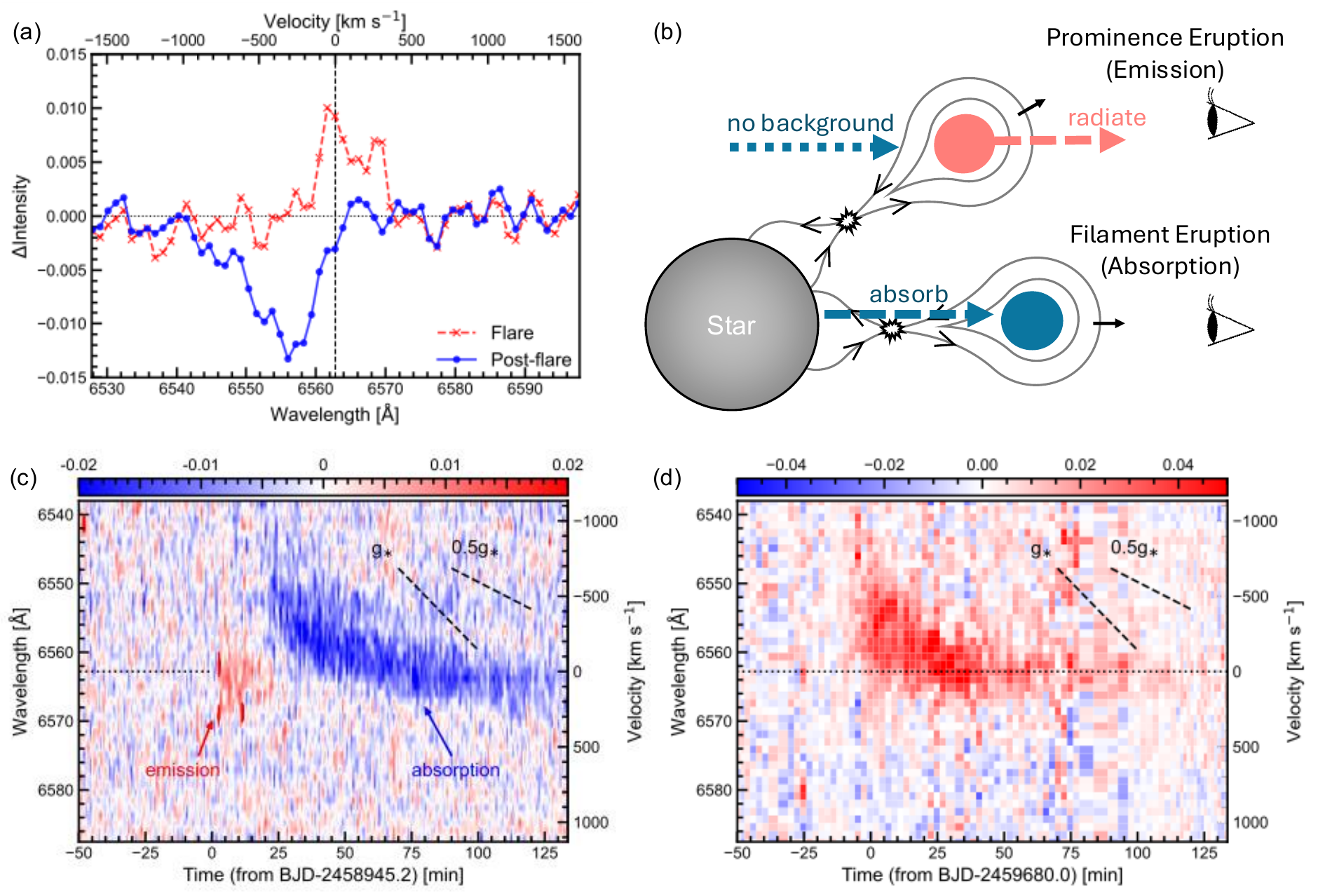}
\caption{
Discovery of filament and prominence eruption signatures from EK Draconis.
(a) Pre-flare-subtracted H$\alpha$ spectra during the flare and post-flare phases, showing a blue-shifted absorption component.
(b) Schematic illustrations of the geometrical origin of blue-shifted absorption and emission signatures. Cool material moving toward the observer in front of the stellar disk produces an absorption component, while material erupting above the stellar limb produces an emission component.
(c,d) Dynamic spectra of the H$\alpha$ line for the blue-shifted absorption event and blue-shifted emission event, respectively. Panels (a,c) are reproduced from \citet{2022NatAs...6..241N}.
Panel (d) is reproduced from \citet{2024ApJ...961...23N}.
}
    \label{fig:filament}
\end{figure}

Young solar-type stars provide a useful regime for testing the solar-stellar connection. Their superflares occur frequently enough to make dedicated monitoring feasible (e.g., once in a few days), while their stellar parameters remain close enough to solar values for solar analogies to be physically meaningful. 
To search for CME-related signatures from such stars, \cite{2022NatAs...6..241N,2022ApJ...926L...5N,2024ApJ...961...23N,2024ApJ...976..255N,2025ApJ...993...80N,namekata2025natas} and \cite{2024MNRAS.532.1486L} conducted a long-term spectroscopic monitoring program.
\cite{2022NatAs...6..241N,2022ApJ...926L...5N,2024ApJ...961...23N,2024ApJ...976..255N,2025ApJ...993...80N,namekata2025natas}  used the low-resolution (R$\sim$2000) spectrograph KOOLS-IFU \citep{2019PASJ...71..102M} on the 3.8 m Seimei Telescope in Okayama, Japan \citep{2020PASJ...72...48K}. 
They selected two nearby young solar-type stars, EK Draconis (G1.5V; age 50--125 Myr; $P_{\rm rot}=2.76$ days; distance 34.4 pc) and V889 Her (G0V; age 30 Myr; $P_{\rm rot}=1.33$ days; distance 35.4 pc), as suitable targets for flare monitoring based on the TESS observations \citep{2025ApJ...993...80N}.
The primary strategy was to detect transient H$\alpha$ line asymmetries as signatures of cool erupting material. 
These intensive campaigns detected fifteen H$\alpha$ superflares, plus one Carrington-class solar-like flare discussed in Section~\ref{sec:multi_wavelength}, providing a new observational basis for studying filament and prominence eruptions from solar-type stars through comparisons with solar events and theoretical models.

The first key event was detected on EK Draconis on 2020 April 5, during a superflare with an energy of approximately $2 \times 10^{33}$ erg \citep{2022NatAs...6..241N}. As shown in Figure~\ref{fig:filament}(a), the post-flare H$\alpha$ spectra exhibited a clear blue-shifted absorption component with a velocity of up to $510~{\rm km~s^{-1}}$. The corresponding dynamic spectrum in Figure~\ref{fig:filament}(c) shows that this absorption feature appeared transiently in the blue wing of H$\alpha$. As illustrated in Figure~\ref{fig:filament}(b), such a blue-shifted absorption is naturally expected when cool material moves toward the observer in front of the stellar disk. The deceleration of the blue-shifted component was also consistent with the surface gravity of EK Draconis, suggesting that the material was ejected from near the stellar surface. Together with the Sun-as-a-star comparison discussed in Section~\ref{sec:sun_as_a_star}, this event provides strong evidence for a stellar filament eruption associated with a superflare. In contrast to many M-dwarf blue-asymmetry events, whose interpretation can be affected by flare-related chromospheric flows, this event is one of the clearest cases of a stellar filament eruption.
Complementarily, \citet{2024MNRAS.532.1486L} reported partial observations of the same event after the strong blueshift phase and found no dominant redshifted component afterward.

A second important event on 2022 April 10 showed a different but equally remarkable observational appearance. During a superflare with an energy of approximately $1.5 \times 10^{33}$ erg, the H$\alpha$ line exhibited a blue-shifted emission component with velocities of about $400$--$690~{\rm km~s^{-1}}$ \citep{2024ApJ...961...23N}. The dynamic spectrum in Figure~\ref{fig:filament}(d) shows the time evolution of this emission feature. Although this event appeared in emission rather than absorption, its temporal evolution was strikingly similar to that of the 2020 absorption event. Its high velocity, reaching $690~{\rm km~s^{-1}}$, and the weak enhancement near the H$\alpha$ line center suggest that the feature is unlikely to be explained simply by flare-related line broadening or chromospheric flows. Instead, the blue-shifted emission component is naturally interpreted as cool material erupting above the stellar limb, as illustrated in Figure~\ref{fig:filament}(b). This event therefore provides evidence for an off-limb prominence eruption from a young solar-type star.

These two contrasting events demonstrate that both absorption and emission signatures can be used to probe cool erupting material from unresolved stellar spectra. In the solar analogy, they correspond to on-disk filament eruptions and off-limb prominence eruptions, respectively. 

\Add{
A closely related observational phenomenon is that of slingshot prominences on rapidly rotating young stars. These cool, magnetically supported condensations are observed in H$\alpha$ absorption while transiting the stellar disk and can appear in emission when seen beyond the stellar limb, showing the same basic dependence on viewing geometry as the filament and prominence signatures described above \citep{1989MNRAS.236...57C,2019MNRAS.482.2853J,2020MNRAS.491.4076J}. Although these structures are not necessarily flare-driven and may remain quasi-stable, their observational signatures overlap with those used in stellar CME searches, motivating closer interaction between the two communities in future studies.}

The following sections discuss how the interpretations of the EK Draconis events are supported by comparisons with solar events and theoretical models, and how their inferred velocities and masses indicate that these stellar eruptions are much faster and more massive than typical present-day solar filament and prominence eruptions.

\section{Solar-stellar connection: Sun-as-a-star analysis}
\label{sec:sun_as_a_star}

\begin{figure}
    \centering
    \includegraphics[width=1.0\linewidth]{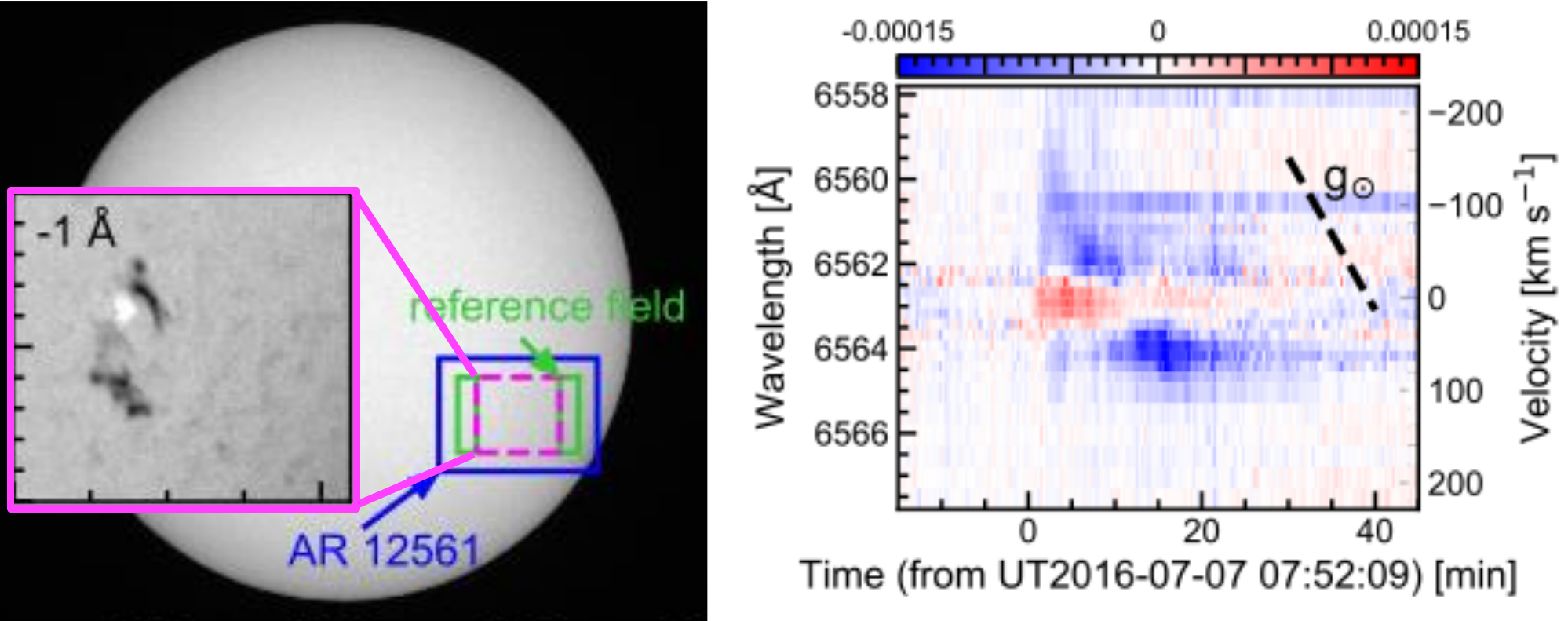}
\caption{
Left: Solar continuum image and an H$\alpha$ filament eruption observed by SMART at Hida Observatory. Right: Sun-as-a-star dynamic spectrum of the H$\alpha$ line for the solar filament eruption. Adapted from \citet{2022NatAs...6..241N}.
}
    \label{fig:sun_as_a_star}
\end{figure}

The interpretation of stellar H$\alpha$ line asymmetries relies strongly on solar comparison. Since solar filament and prominence eruptions can be spatially resolved and directly linked to CMEs, the concept of the ``Sun-as-a-star" analysis provides an essential calibration for asking how stellar CME signatures should appear in unresolved spectra and for evaluating possible false positives.

\citet{2022NatAs...6..241N} analyzed solar filament eruption and surge events, likely associated with CMEs \citep[cf.][]{2019SunGe..14...95S}, and performed Sun-as-a-star H$\alpha$ analysis. Figure~\ref{fig:sun_as_a_star} illustrates this approach for a solar filament eruption. Remarkably, the disk-integrated H$\alpha$ signature closely resembles the stellar event shown in Figure~\ref{fig:filament}. As in the EK Draconis event, the solar H$\alpha$ dynamic spectrum shows a strong blue-shifted absorption component in the early phase, followed by gradual deceleration. The light curve also shows a post-flare dimming-like feature after the flare emission, similar to the stellar case. These early solar--stellar comparisons provided important support for interpreting the EK Draconis event as a solar-like, but much larger, filament eruption.

Recently, this approach has been extended to a wider variety of solar events, allowing the diversity of disk-integrated H$\alpha$ signatures to be examined more systematically \citep[e.g.,][]{2022ApJ...933..209N,2022ApJ...939...98O,2024ApJ...974L..13O,2024ApJ...964...75O,2024A&A...682A..46P,2024ApJ...966...45M,2025A&A...700A.275D,2025ApJ...993..126L}.  
Sun-as-a-star analyses are not limited to H$\alpha$, and similar studies have also examined how CMEs appear as EUV dimming or Doppler-shifted signatures in spatially unresolved observations \citep{2016SoPh..291.1761H,2021NatAs...5..697V,2022ApJ...931...76X,2025ApJ...988..167M}. 
More recently, multi-wavelength Sun-as-a-star analyses of individual solar events have begun to connect these different diagnostics \citep{2024ApJ...964...75O}.

These studies are essential for interpreting the recent multi-wavelength searches for stellar CME signatures discussed in Section~\ref{sec:multi_wavelength}. 
Further Sun-as-a-star studies will still be necessary to bridge the gap between solar and stellar CME diagnostics, as discussed in Section~\ref{sec:future}.

\section{Data-driven modeling}\label{sec:data_driven_modeling}

H$\alpha$ observations alone cannot determine the overall structure and fate of stellar eruptions. 
Numerical and forward modeling are therefore needed to connect the observed spectral signatures with eruption geometry, magnetic expansion, and possible CME escape.

A useful first step is to compare the observed stellar eruptions with solar-based models. \citet{2024ApJ...963...50I} developed a pseudo two-dimensional MHD model for the filament eruption detected in the 2020 EK Draconis superflare \citep{2022NatAs...6..241N}, showing that the observed blue-shifted H$\alpha$ absorption can be explained as cool erupting material embedded in an expanding magnetic structure with a constant velocity of $\sim$500 km s$^{-1}$, i.e., a successful escaping CME case. This result supports the idea that the same basic physical framework used for solar filament eruptions can be applied to young solar-type stars.

\begin{figure}
    \centering
    \includegraphics[width=0.9\linewidth]{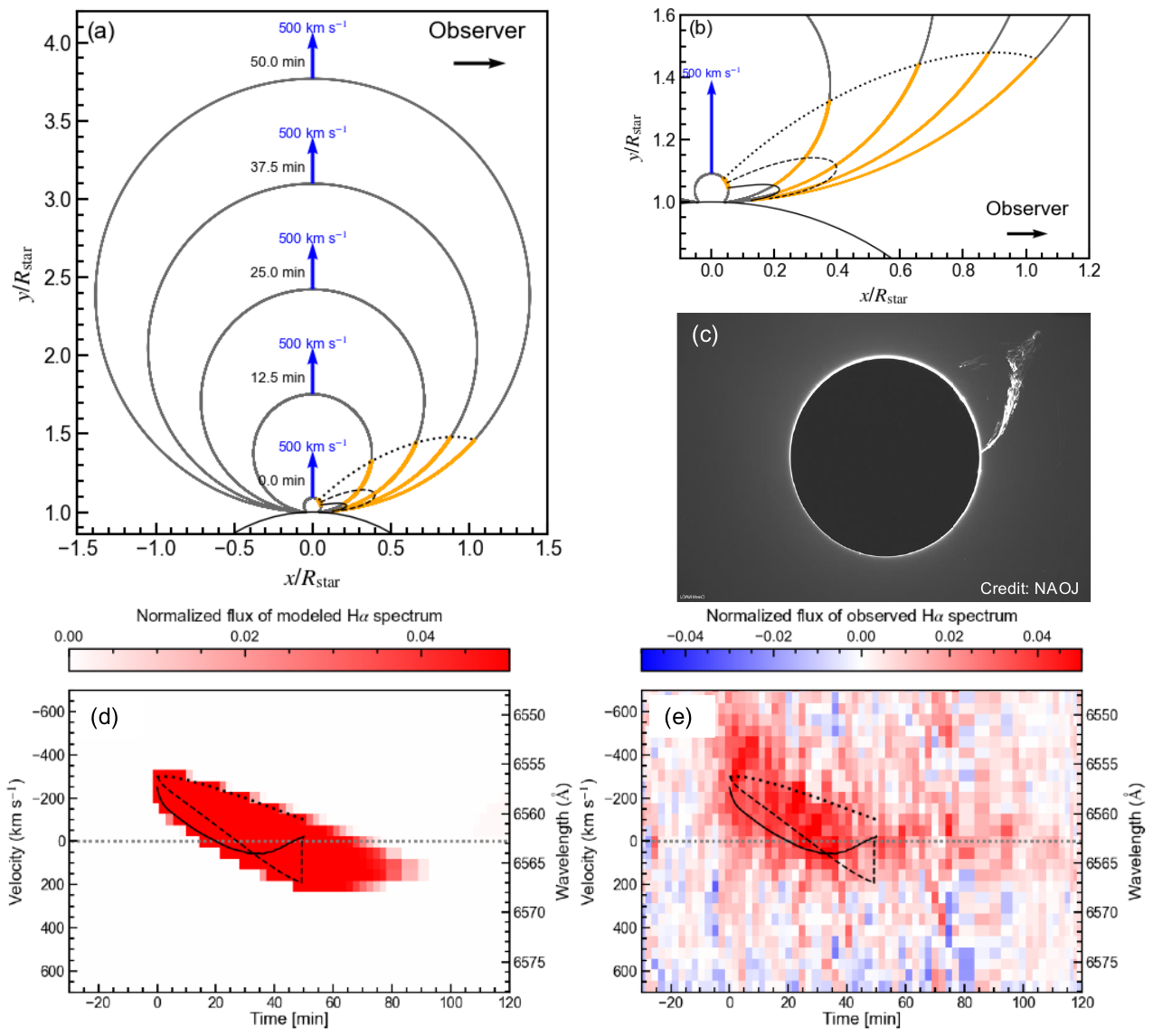}
    \caption{
    Data-driven modeling of the 2022 April 10 prominence eruption on EK Draconis, adapted from Figures 11 and 12 of \citet{2024ApJ...976..255N}. 
    (a,b) Geometry assumed in the pseudo two-dimensional MHD model by \cite{2024ApJ...963...50I}, in which prominence material is placed along an expanding magnetic loop whose footpoints are located near the stellar limb. The orange structure represents the prominence material, and the black lines indicate Lagrangian trajectories of plasma elements.
    (c) A coronagraph at the Norikura Solar Observatory photographed a giant erupting prominence on July 31, 1992 (Credit: NAOJ; \url{https://www.nao.ac.jp/en/gallery/weekly/2018/20180213-prominence.html}). For a better comparison with the model, the image is horizontally flipped.
    (d,e) Comparison between the simulated and observed H$\alpha$ dynamic spectra. The simulated spectrum reproduces the main time evolution of the blue-shifted emission component, supporting the interpretation that the observed H$\alpha$ feature traces the cool prominence material embedded in a larger expanding magnetic structure. 
    }
    \label{fig:ekdra_data_driven_model}
\end{figure}

For the 2022 April 10 prominence eruption, the modeling was further developed by combining a simple trajectory model and a pseudo two-dimensional MHD calculation \citep{2024ApJ...976..255N}. First, a one-dimensional free-fall model was used to constrain the geometry of the prominence motion from the observed velocity evolution. The preferred solution suggested that the prominence was launched from near the stellar limb with an angle of roughly $20^\circ$. This geometry was then adopted in the pseudo two-dimensional model, allowing a more direct comparison between the observed H$\alpha$ dynamic spectrum and the predicted time evolution of the erupting cool material.
As shown in Figure~\ref{fig:ekdra_data_driven_model}, the model successfully reproduced the observed H$\alpha$ dynamic spectrum, including the decelerating blue-shifted emission component. 

An important implication of these models is that the cool prominence material can decelerate under gravity and magnetic forces, while the surrounding magnetic loop continues to expand outward and may evolve into a CME. Such modeling therefore provides an essential bridge between observations and CME physics. 
It also makes it possible to test whether the observed line profiles are consistent with solar-like eruption geometry, to estimate the three-dimensional trajectory of the cool material, and to examine whether the magnetic structure can continue to expand even after the prominence itself begins to decelerate.

\section{Do escaping CMEs occur?}
\label{sec:do_CMEs_occur}

\begin{figure}
    \centering
    \includegraphics[width=0.48\linewidth]{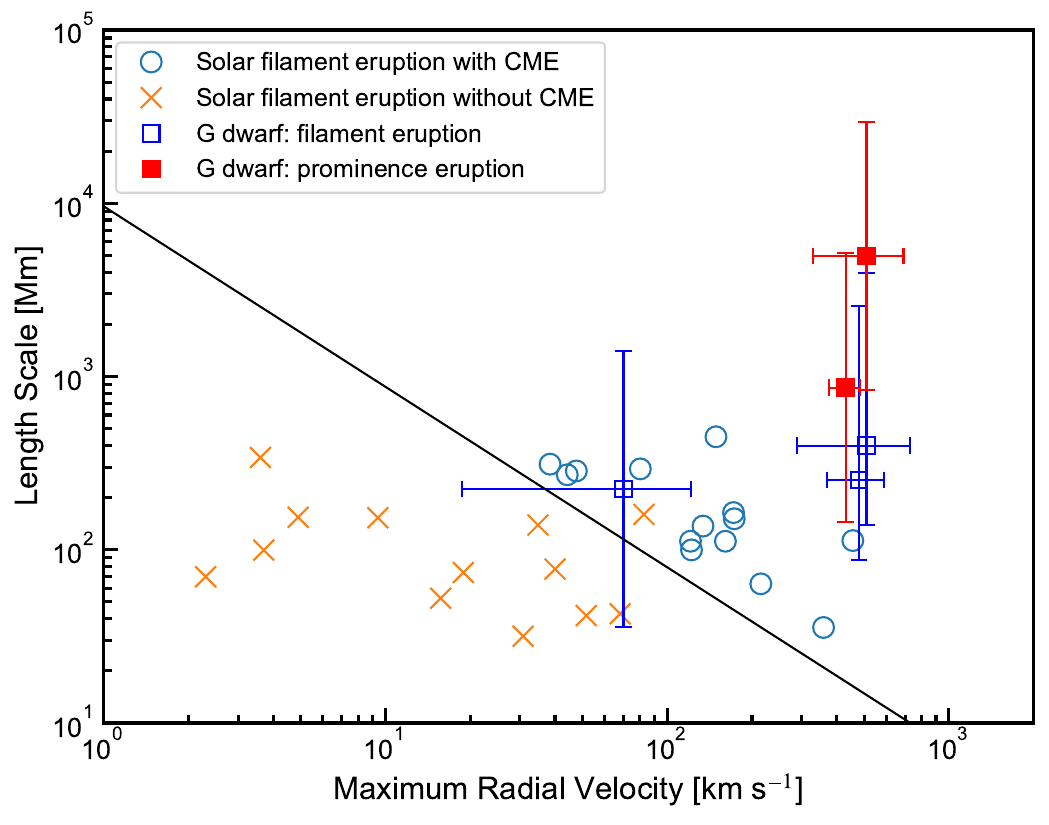}
    \caption{
    Updated view of velocity-length-scale comparison of solar and stellar filament/prominence eruptions. 
    Solar filament eruptions with and without associated CMEs are compared with the EK Draconis events. 
    The empirical solar threshold (solid line) proposed by \citet{2021EP&S...73...58S} roughly separates CME-associated and non-CME filament eruptions. 
    Stellar data are adapted from \citet{2025ApJ...993...80N}, with additional data from \citet{namekata2025natas}. 
    One data point for the 2023 March 8 event is slightly corrected from the original plot, which does not change the conclusion in this order-of-magnitude discussion. Velocity dispersions are consistently shown as velocity uncertainties, while for the 2022 April 10 event, the lower and upper velocities represent two fitting methods examined in \citet{2024ApJ...961...23N}.
    The data from solar-type stars are accessible through \url{https://github.com/KosukeNamekata/publications/tree/main/data/EKDra}.
    }
    \label{fig:vels}
\end{figure}

A key question is whether the observed filament and prominence eruptions actually evolve into escaping CMEs. 
H$\alpha$ observations trace only cool material, which corresponds to the filament or prominence core in the solar analogy. Therefore, the detection of a filament/prominence eruption does not by itself prove that the entire magnetic structure escapes from the star.
Nevertheless, several lines of evidence support the escaping-CME interpretation as follows.

First, the observed velocities are already very large. The EK Draconis events show line-of-sight velocities of up to $690~{\rm km~s^{-1}}$, comparable to the stellar escape velocity and larger than those of typical solar filament eruptions associated with CMEs, which are usually a few hundred ${\rm km~s^{-1}}$. Since these are projected velocities, the true velocities can be larger. In addition, the velocity dispersion of the H$\alpha$ components suggests top speeds of $\sim 660$--$1080~{\rm km~s^{-1}}$, exceeding the escape velocity of EK Draconis \citep{2024ApJ...961...23N,2025ApJ...993...80N}. Solar observations also show that the CME front propagates several times faster than the associated filament or prominence \citep{2003ApJ...586..562G}, implying that the actual CME front speed could reach a few times $1000~{\rm km~s^{-1}}$ or more. Thus, a mean H$\alpha$ velocity below the surface escape velocity does not necessarily imply a failed eruption; rather, the observations are consistent with fast CMEs.

Second, the velocity-length-scale relation provides an empirical solar criterion. \citet{2021EP&S...73...58S} showed that solar filament eruptions with larger velocity-length scales, which can be regarded as an empirical proxy for eruption scale or kinematic strength, are more likely to be associated with CMEs. As shown in Figure~\ref{fig:vels}, the EK Draconis filament and prominence eruptions lie well above the empirical threshold for CME-associated solar events \citep{2025ApJ...993...80N}. This suggests that these stellar eruptions have sufficiently large momentum or kinetic scale to evolve into CMEs.

Third, data-driven modeling supports the same picture. The pseudo two-dimensional MHD models discussed in Section~\ref{sec:data_driven_modeling} show that the cool prominence material can decelerate, while the surrounding magnetic loop continues to expand outward and may evolve into a CME \citep{2024ApJ...963...50I,2024ApJ...976..255N}. This is important because H$\alpha$ traces only the cool component. A decelerating or sub-escape H$\alpha$ feature therefore does not necessarily mean that the entire eruption is confined.

Taken together, these arguments lead to the first conclusion of this review: the detected EK Draconis filament and prominence eruptions do not provide direct images of escaping CMEs, but they provide strong indirect evidence that at least some superflares on young solar-type stars are accompanied by escaping CMEs. More robust confirmation will require additional diagnostics, including coronal dimming, UV line shifts, and radio signatures. These multi-wavelength constraints are discussed in Section~\ref{sec:multi_wavelength}.

\section{Mass scaling and kinetic-energy controversy}\label{sec:velocity_mass_kinetic_energy}

\begin{figure}
    \centering
    \includegraphics[width=0.85\linewidth]{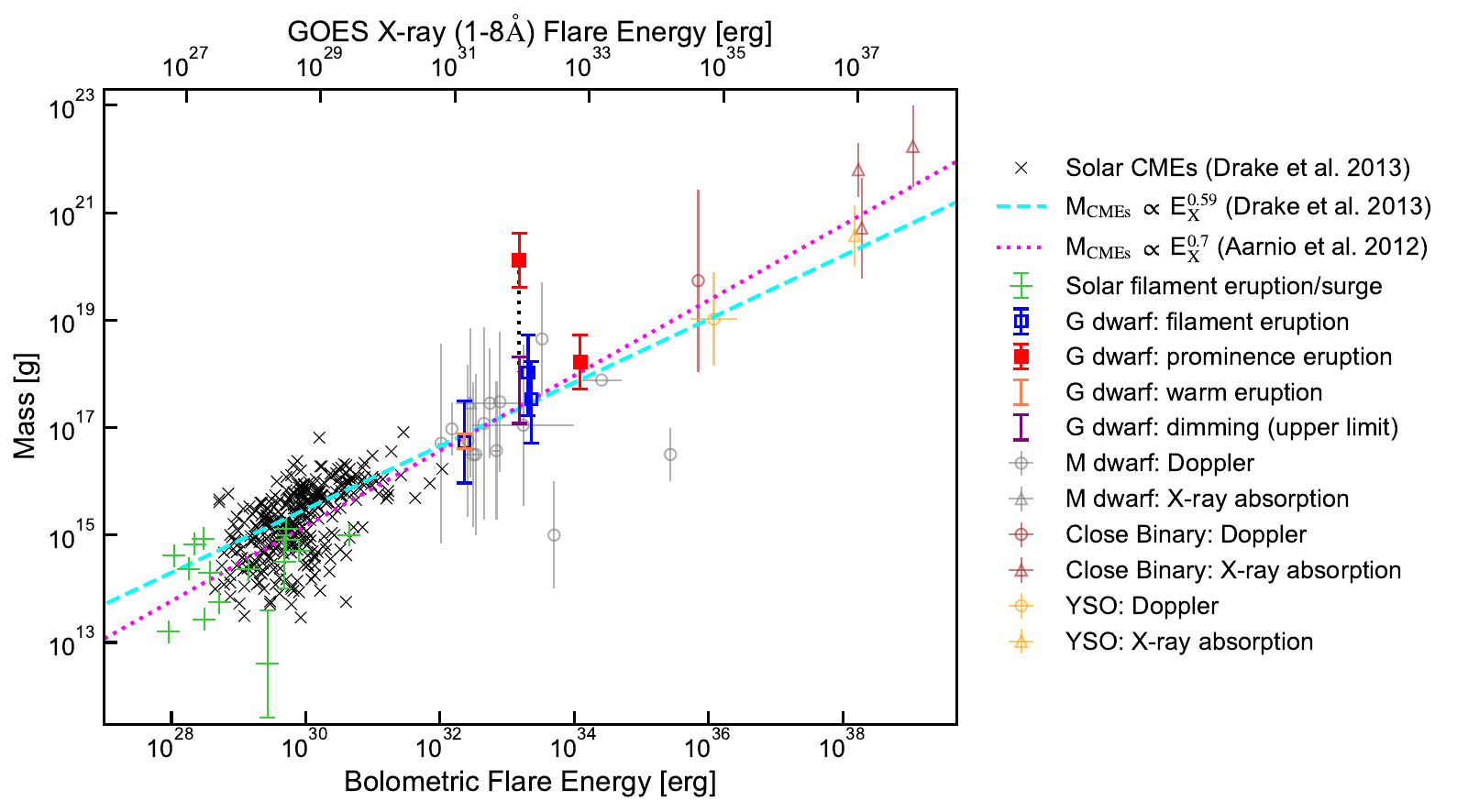}
    \includegraphics[width=0.85\linewidth]{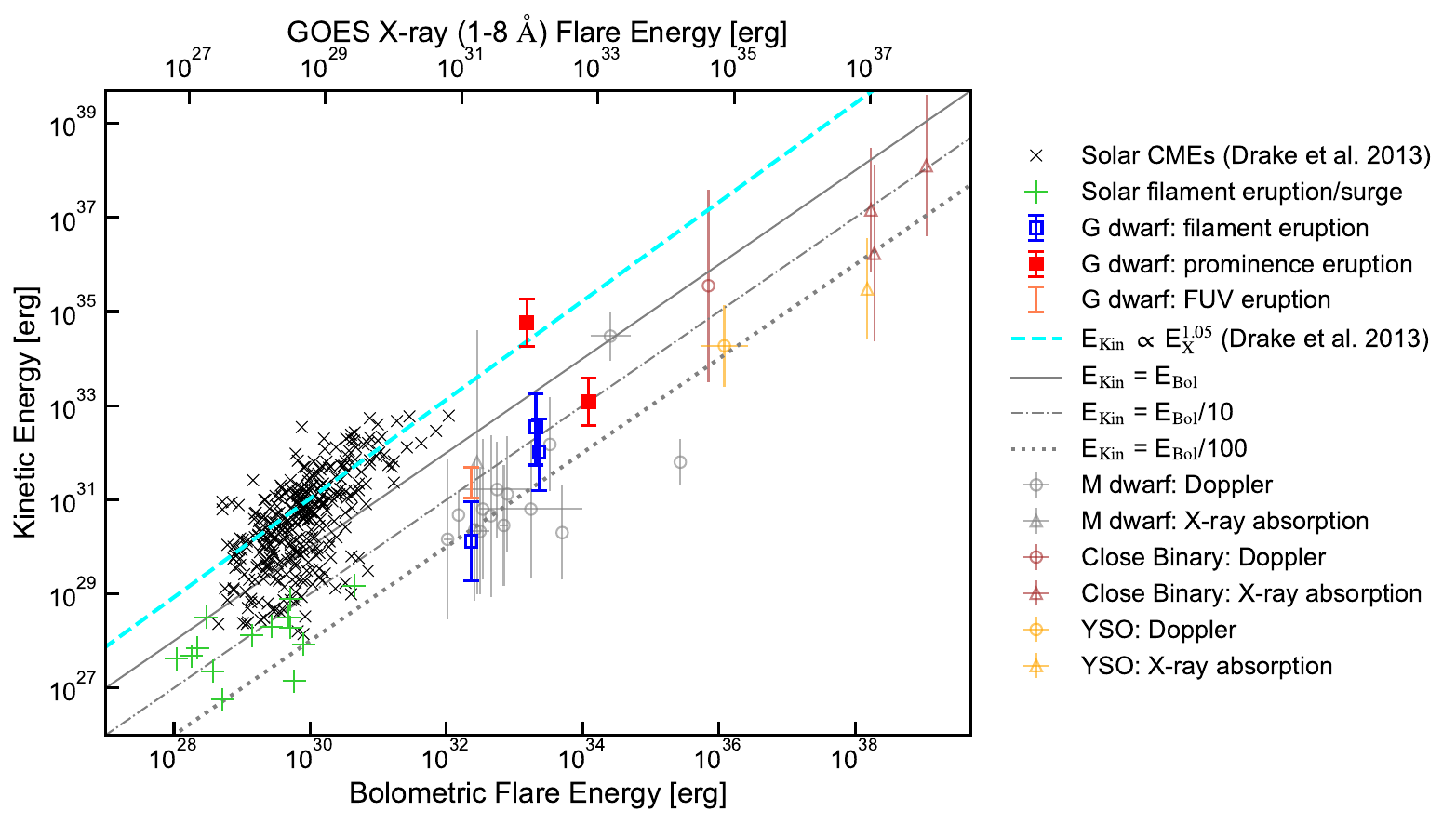}
    \caption{
    An updated view of physical properties of solar and stellar filament/prominence eruptions and CMEs.
    Upper: Erupting mass as a function of flare energy.
    Lower: Kinetic energy as a function of flare energy.
    The stellar types include the Sun \citep{2013ApJ...764..170D,2019SunGe..14...95S,2022NatAs...6..241N,2023ApJ...943..143K}, G dwarf \citep{2022NatAs...6..241N,2024ApJ...961...23N,2025ApJ...993...80N,namekata2025natas}, M dwarfs (\citealt{2019ApJ...877..105M} and references therein, note that AT Mic is an M-dwarf binary; \citealt{2020PASJ..tmp..253M,2024ApJ...961..189N,2024PASJ...76..175I}), close binaries (active close binary, interacting or evolved binary; \citealt{2019ApJ...877..105M} and references therein; \citealt{2023ApJ...948....9I}), and a young stellar object (YSO; \citealt{2019ApJ...877..105M} and references therein).
    The values were taken from studies that carefully defined the relevant parameters and flare energy ranges.
    The empirical solar relations derived by \cite{2013ApJ...764..170D} and \cite{2012ApJ...760....9A} are plotted.
    For putting all solar and stellar data together, GOES-band X-ray flare energies were converted to bolometric flare energies by adopting a single energy partition, 
    $E_{\rm X,GOES}=0.06E_{\rm bol}$ \citep{2015ApJ...809...79O}.
    For data or relations reported only in terms of the GOES peak flux (flare class), we used $E_{\rm bol}\,[{\rm erg}]=10^{35}F_{\rm GOES}\,[{\rm W\,m^{-2}}]$.
    Some previous solar studies adopted an empirical factor of $E_{\rm X,GOES}\sim0.01E_{\rm bol}$; therefore, comparisons with those studies may include a systematic offset but this uncertainty is not critical for the present order-of-magnitude comparison.
    The data from solar-type stars are accessible through \url{https://github.com/KosukeNamekata/publications/tree/main/data/EKDra}.
    }
    \label{fig:statistics}
\end{figure}

The EK Draconis eruptions are remarkable not only in velocity, but also in mass and kinetic energy. The inferred masses of the cool filament/prominence material reach $\sim 10^{17}$--$10^{20}$ g, with the largest event reaching $\sim 10^{19}$--$10^{20}$ g \citep{2024ApJ...961...23N,2025ApJ...993...80N}. These values are much larger than typical present-day solar filament/prominence masses, and the corresponding kinetic energies also lie at the high-energy end of solar and stellar eruptive phenomena. In this sense, the detected events are much more massive and energetic versions of solar filament/prominence eruptions.
Figure~\ref{fig:statistics} summarizes updated relations between flare energy and erupting mass or kinetic energy, including events from the Sun, solar-type stars, M dwarfs, close binaries, and a young stellar object (YSO).
This comparison provides a useful way to examine similarities and differences among stellar types and observational signatures within a common framework.

As shown in Figure~\ref{fig:statistics}, the inferred eruption masses for solar-type stars are broadly consistent with extrapolations of solar CME/eruption mass--flare energy relations \citep{2012ApJ...760....9A,2013ApJ...764..170D,2016ApJ...833L...8T,2023ApJ...943..143K}. 
A similar relation is also seen for CME candidates from M dwarfs, close binary stars, and YSOs \citep[see][for original plots]{2017ApJ...850..191M}.
Different observational signatures, including coronal dimming, X-ray absorption, and FUV blueshifts, are also included here (see Sections~\ref{sec:observational_diagnostics} and \ref{sec:multi_wavelength}). Despite the different diagnostics and assumptions involved, they are broadly consistent with the overall trend.
This suggests that stellar superflares and associated eruptions may be energized by the same magnetic energy release process that links solar flares and CMEs/eruptions \citep[see physical explanations from][]{2016ApJ...833L...8T,2023ApJ...943..143K}.

The kinetic energy is more controversial and requires a more careful comparison. Extrapolations of solar flare-CME relations \citep{2013ApJ...764..170D,2012ApJ...760....9A} suggested that superflares could produce extremely energetic CMEs, whereas later compilations of stellar CME candidates found that their kinetic energies often appear lower than these naive solar extrapolations \citep{2017ApJ...850..191M,2019ApJ...877..105M}. The EK Draconis events also fall below some extrapolated solar CME kinetic-energy relations, as shown in Figure~\ref{fig:statistics}. 
This apparent kinetic-energy deficit is an important but still controversial issue. One interpretation is physical: strong overlying magnetic fields on active stars may suppress or decelerate eruptions, reducing the kinetic energy of escaping CMEs \citep{2013ApJ...764..170D,2018ApJ...862...93A}. Another interpretation is observational \citep[e.g.,][]{2022NatAs...6..241N,2024ApJ...961...23N}. 
The kinetic energies inferred from H$\alpha$ and other Balmer lines (and even X-ray absorption) are based only on the cool filament/prominence core, whose velocity is projected along the line of sight and, by solar analogy, is expected to be slower than the surrounding coronal CME front \citep[typically 2--8 times;][]{2003ApJ...586..562G}. Thus, the observed kinetic energies should be regarded as lower limits, and a low H$\alpha$-based value does not necessarily imply a weak or confined CME. 
Indeed, solar filament eruptions also lie well below the simple CME kinetic-energy scaling. When solar and stellar filament/prominence eruptions are compared together, they appear to follow a broadly common trend with $E_{\rm kin}/E_{\rm bol}\sim 0.01-1$.
Resolving this issue requires multi-wavelength diagnostics that can trace the hotter and faster components of the eruption.

\section{Frequency and CME-driven mass loss}\label{sec:frequency}

A key question is how often superflare-associated eruptions occur. This is important because their impact on stellar evolution and planetary environments depends on occurrence rate (Section \ref{sec:implication}). In this context, \cite{2025ApJ...993...80N} performed a five-year dedicated H$\alpha$ monitoring program of EK Draconis and V889 Her.
The campaign detected 15 H$\alpha$ superflares, including two blue-shifted absorption events and two blue-shifted emission events interpreted as filament and prominence eruptions, respectively\footnote{This result is based only on superflares and does not include the Carrington-class flare reported by \citet{namekata2025natas}, in order to keep a consistent energy threshold of $10^{33}$ erg.}. Based on these events, the lower limit of the eruption--flare association rate is estimated to be $27^{+25}_{-16}\%$. The corresponding lower-limit occurrence frequency is $0.21 \pm 0.12$ events day$^{-1}$ for EK Draconis, while the upper limit for V889 Her is $<0.32^{+0.46}_{-0.32}$ events day$^{-1}$. These values provide direct observational estimate of the frequency of super-CME-related filament/prominence eruptions from young solar-type stars.
This is broadly consistent with the other campaign by \cite{2024MNRAS.532.1486L}.

These rates should be regarded as lower limits because H$\alpha$ eruptions can be missed owing to projection effects, limited time coverage, low signal-to-noise ratio, and unfavorable geometry. Simple visibility modeling of in-disk filaments gives detection probabilities of only $\sim 10$--$30\%$ for plausible initial velocities \citep{2025ApJ...993...80N}. Thus, the small number of detected blue-shifted events does not necessarily imply a low intrinsic eruption rate, but rather reflects the limited visibility of superflare-associated eruptions in the Doppler-shift method.

Comparison with other stellar types is also useful, although current samples remain small. 
For M dwarfs, recent flare surveys have reported eruption-candidate association rates of order $\sim 15\%$ \citep{2024ApJ...961..189N,2025ApJ...979...93K}, broadly comparable to the lower limit inferred here for young solar-type stars. 
Any difference between G and M dwarfs is therefore not yet statistically significant, but may reflect differences in magnetic topology \citep[e.g., stronger and more dipole-dominated large-scale fields in some low-mass stars;][]{2021A&ARv..29....1K} or in the visibility of filament/prominence signatures \citep{2022MNRAS.513.6058L}.

Combining the occurrence rate with estimated eruptive masses gives a lower limit on the CME-driven mass-loss rate. For EK Draconis, the inferred lower limit is of order $4 \times (10^{-13}$--$10^{-12})~M_\odot~{\rm yr^{-1}}$ \citep{2025ApJ...993...80N}. This estimate assumes that super-CMEs carry masses comparable to the observed filament/prominence eruptions, which is motivated by solar comparisons between CME masses and filament/prominence masses \citep{2012ApJ...760....9A,2013ApJ...764..170D,2023ApJ...943..143K,2024ApJ...961...23N}. 
The resulting value is broadly consistent with the indirect estimate by \citet{2015ApJ...809...79O}, who inferred a CME mass-loss rate of order $9 \times (10^{-13}$--$10^{-12})~M_\odot~{\rm yr^{-1}}$ for EK Draconis from X-ray flare statistics and solar CME scaling relations. 
\Add{The lower-limit estimate for EK Draconis is also comparable to the lower end of the $\sim10^{-12}$--$10^{-11}\,M_\odot\,\mathrm{yr}^{-1}$ range predicted for a Sun-like star by the empirical model of \citet{2017MNRAS.472..876O}, although the actual rate may be lower than such solar-based extrapolations if CMEs are more strongly confined on active stars.}
Even this lower limit is comparable to the expected steady wind mass-loss rate of young solar-type stars \citep{2011ApJ...741...54C,2021LRSP...18....3V,2023ApJ...957...71S}, suggesting that transient CME-driven mass loss may be important for stellar mass and angular momentum evolution.


\section{Magnetic field environment}\label{sec:magnetic}

Measurements of stellar magnetic-field topology are crucial for constraining the possible source regions of eruptions and for understanding whether they escape or are confined by the surrounding magnetic field \citep{2018ApJ...862...93A,2022MNRAS.509.5075S,2024MNRAS.533.1156S}. 
Although simultaneous observations of eruptive signatures and magnetic-field topology are still rare, recent studies have begun to pursue this direction.


\begin{figure}
    \centering
    \includegraphics[width=0.48\linewidth]{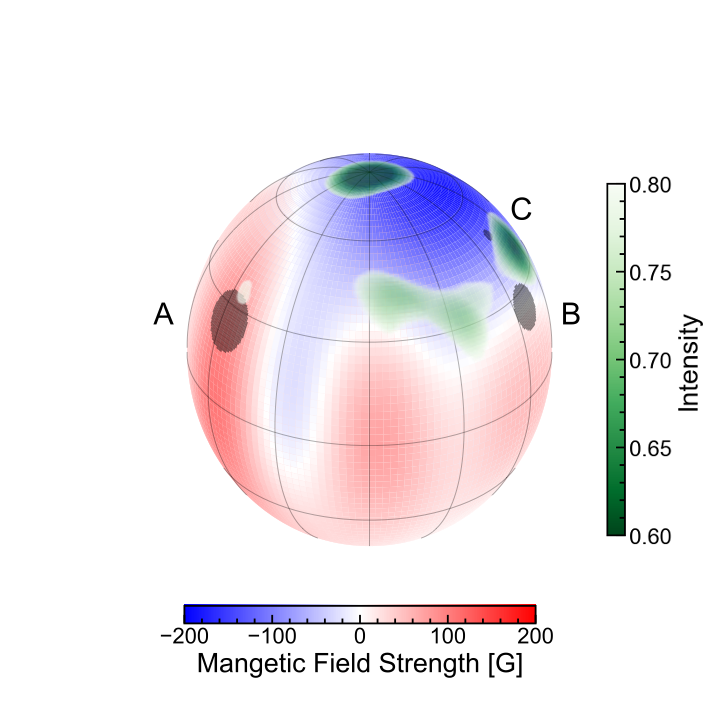}
    \includegraphics[width=0.48\linewidth]{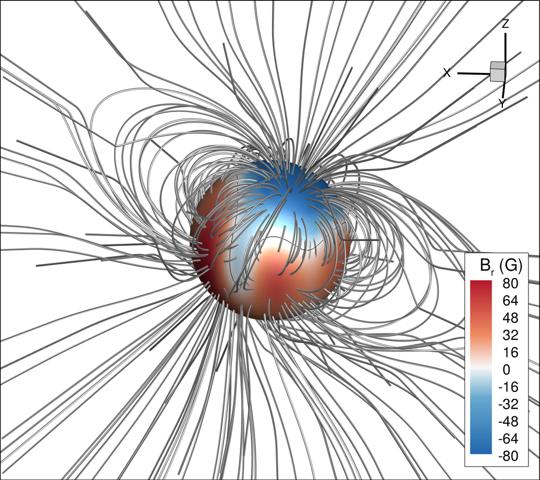}
    \caption{
    Magnetic environment of EK Draconis at the time of the 2022 April 10 prominence eruption. 
    Left: comparison of the Doppler Imaging spots, TESS light-curve-modeling spots, and radial magnetic field map, adapted from Figure 5(c) of \citet{2024ApJ...976..255N}. 
    Right: potential-field extrapolation of the radial magnetic field, adapted from Figure 4 of \citet{2024ApJ...976..255N}. 
    }
    \label{fig:ekdra_magnetic_environment}
\end{figure}

For the 2022 April 10 prominence eruption on EK Draconis, the magnetic environment was investigated using simultaneous TESS light-curve inversion, Doppler Imaging, and Zeeman Doppler Imaging \citep{2024ApJ...976..255N,2026ApJ..1001...18I}. 
Figure~\ref{fig:ekdra_magnetic_environment} shows the magnetic field and intensity maps when the prominence eruption occurred. The TESS light-curve inversion identified three mid-latitude spots visible around the time of the eruption, while the DI/ZDI map showed a polar spot. In particular, Spot B is located close to a polarity inversion line, and Spot A may also be associated with such a region within the spatial resolution of the mapping. Since solar prominences often form above polarity inversion lines, these mid-latitude spots could be plausible source regions for the stellar prominence eruption, although this is in the closed coronal loop and may need a more careful consideration.
Another possibility is that the eruption originated from, or escaped through, the high-latitude region connected to more open magnetic fields. Thus, both the mid-latitude PIL regions and high-latitude open-field regions remain physically plausible.

\begin{figure}
    \centering
    \includegraphics[width=0.45\linewidth]{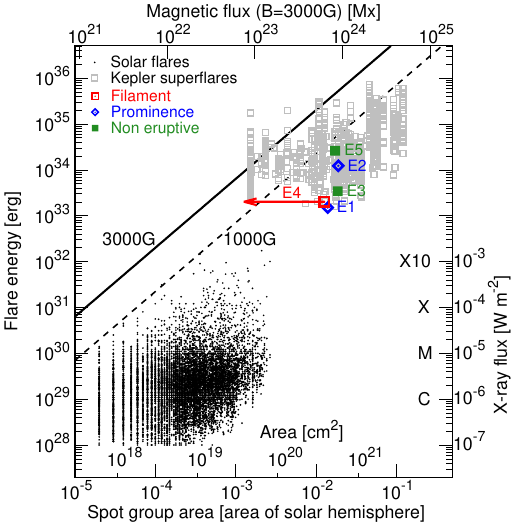}
    \includegraphics[width=0.45\linewidth]{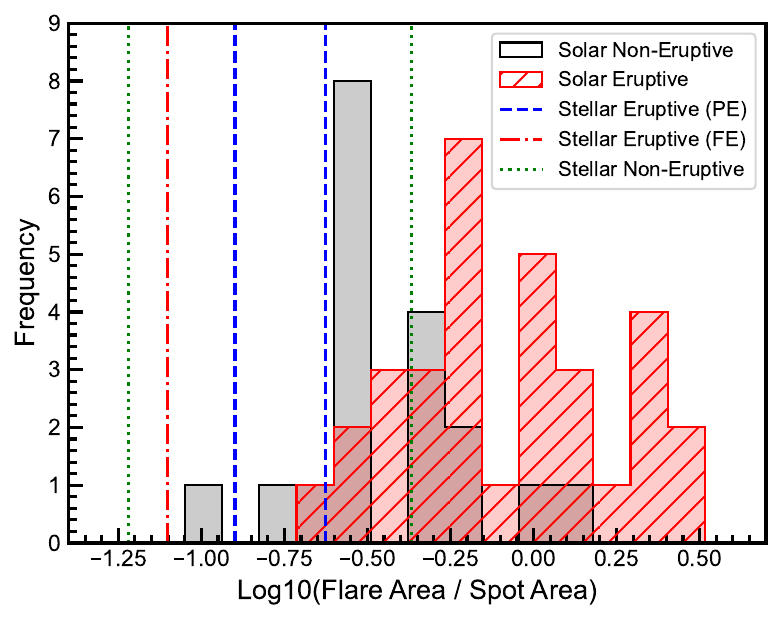}
    \caption{
    Comparison between flare scale and spot scale, adapted from Figure 13 of \citet{2024ApJ...976..255N}. 
    The figure compares EK Draconis superflares with solar flares and Kepler solar-type-star superflares in terms of flare energy relative to the estimated spot magnetic energy and flaring area relative to spot area. 
    These quantities provide solar-motivated diagnostics for examining whether a flare is likely to be eruptive or confined.
    }
    \label{fig:ekdra_flare_spot_relation}
\end{figure}

A related diagnostic is the comparison between flare scale and spot scale. Solar studies suggest that flares whose ribbon or energy scale is large relative to the host active-region scale are more likely to be eruptive than confined \citep{2017ApJ...834...56T,2023ApJ...958..104K}. Motivated by this idea, \citet{2024ApJ...976..255N} compared EK Draconis flares with the flare energy--spot area relation \citep{2020arXiv201102117O} and with the ratio of flaring area to spot area \citep{2017ApJ...850...39T}, as shown in Figure~\ref{fig:ekdra_flare_spot_relation}. The current EK Draconis sample is still too small to show a clear separation between eruptive and non-eruptive superflares. 
Nevertheless, this initial study suggests a useful way to connect solar and stellar eruptive-flare diagnostics.
Future progress will require systematic observations of both eruptions and their magnetic-field environments, so that the relation among starspots, magnetic topology, and CME escape can be tested statistically.

\section{Multi-wavelength diagnostics}\label{sec:multi_wavelength}

The discussion so far has focused mainly on H$\alpha$ signatures of cool filament/prominence material. However, as summarized in Figure~\ref{fig:overall}, a CME is intrinsically a multi-temperature and multi-component phenomenon, including a cool filament core, hotter plasma, shocks, and energetic particles.  Multi-wavelength diagnostics are therefore essential for testing whether the H$\alpha$-detected eruptions are part of larger CME systems and for constraining their full physical properties.
Here, two pioneering multi-wavelength studies are introduced in Sections~\ref{sec:10-1} and~\ref{sec:10-2}. 
Section~\ref{sec:10-3} further discusses radio efforts for solar-type stars as a complementary diagnostic of shocks and escaping plasma at larger heights.

\subsection{H$\alpha$ and X-ray diagnostics}\label{sec:10-1}

The 2022 April campaign of EK Draconis demonstrated the value of combining optical and X-ray diagnostics. In the 2022 April 10 event, the blue-shifted H$\alpha$ emission component was interpreted as an off-limb prominence eruption  \citep[see Figure \ref{fig:filament};][]{2024ApJ...961...23N}. Nearly simultaneous NICER observations showed a possible X-ray coronal dimming after the prominence eruption. On the Sun, coronal dimming is interpreted as a decrease in coronal emission measure caused by plasma evacuation during a CME \citep{2016SoPh..291.1761H,2021NatAs...5..697V,2025LRSP...22....2V}. The stellar dimming signal remains tentative, but its timing relative to the H$\alpha$ eruption is suggestive of a CME-related coronal response.

An interesting result is that the mass estimates from H$\alpha$ and X-rays differ substantially, as in Figure \ref{fig:statistics} (see points connected with a black dotted line). If the post-flare X-ray decrease is interpreted as coronal dimming, the evacuated hot coronal mass is estimated to be only $\sim 10^{17}$--$10^{18}$ g, depending on the assumed coronal density, whereas the H$\alpha$ prominence mass is $\sim 4.0\times10^{19}$--$4.2\times10^{20}$ g \citep{2024ApJ...961...23N}. This difference does not necessarily contradict the CME interpretation, because H$\alpha$ and X-rays trace different temperature and spatial regions, although their differences are very large and each estimate has its own large uncertainty. Rather, it highlights that mass estimates from different diagnostics may not be directly interchangeable and require solar calibration and physical modeling.

\subsection{Far-UV and H$\alpha$ diagnostics}\label{sec:10-2}

\begin{figure}
    \centering
    \includegraphics[width=1.0\linewidth]{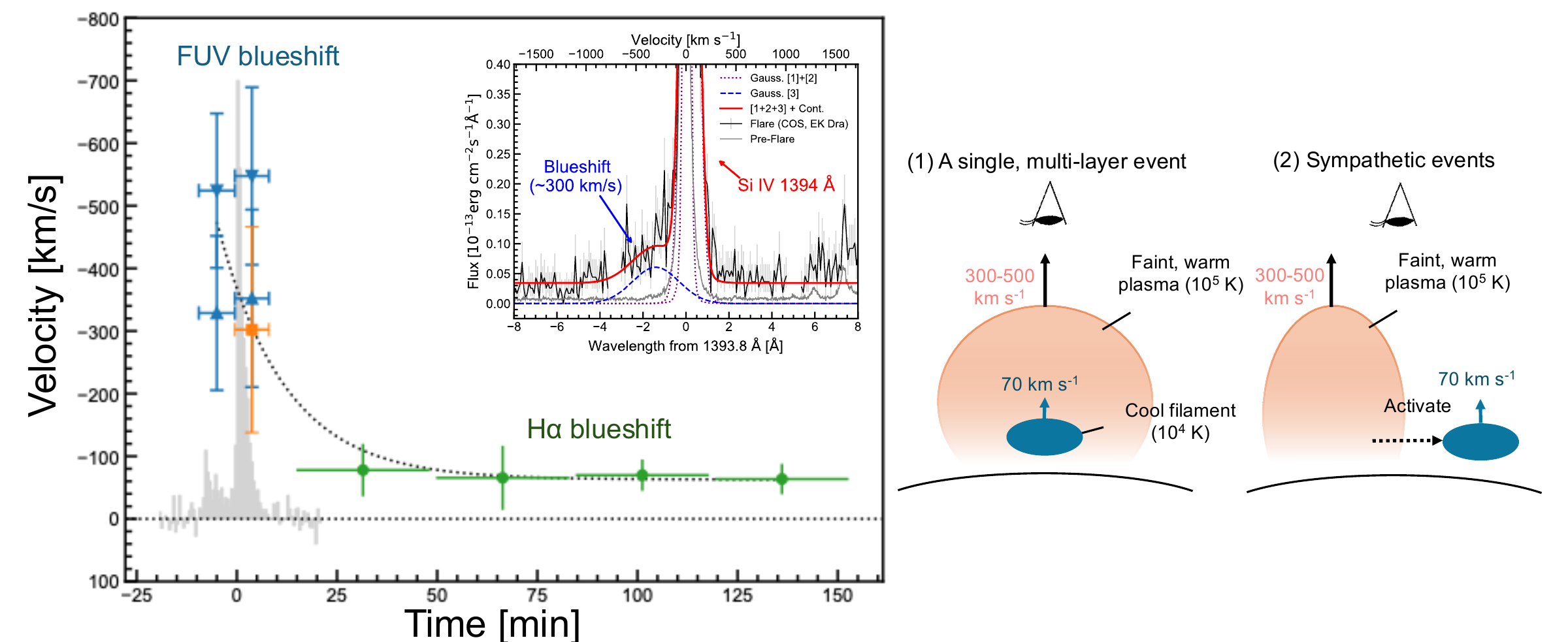}
    \caption{
    Multi-temperature CME signatures from EK Draconis, adapted from \citet{namekata2025natas}. 
    Left: time evolution of the FUV and H$\alpha$ blueshifted components. 
    Right: schematic interpretation in which the fast warm FUV component and the slower cool H$\alpha$ component trace different temperature layers of a single magnetic eruption, or two distinct but physically connected eruptions.
    }
    \label{fig:ekdra_multitemperature}
\end{figure}

UV spectroscopy provides a direct probe of warmer erupting plasma. Coordinated HST/COS and ground-based observations of EK Draconis revealed fast FUV blueshifted components during a Carrington-class flare, with a bolometric energy of $2.3\times10^{32}$ erg \citep{namekata2025natas}. As in Figure \ref{fig:ekdra_multitemperature}, transition-region lines such as C~{\sc iii} and Si~{\sc iv}, formed at $\sim 10^5$ K, showed blueshifted components with velocities of $\sim 300$--$550~{\rm km~s^{-1}}$, and maximum blueshifted velocities reaching up to $\sim 690~{\rm km~s^{-1}}$. About 10 minutes later, a slower and longer-lasting H$\alpha$ blueshift appeared, tracing cool $\sim 10^4$ K material moving at about $60$--$70~{\rm km~s^{-1}}$ for more than 2 hours. This evolution is difficult to explain as a single plasma component simply cooling and decelerating by solar analogy. 
Instead, it suggests either different temperature layers of one eruption or distinct but physically connected eruptions. 
Figure~\ref{fig:ekdra_multitemperature} summarizes this multi-temperature picture, showing that stellar CME-related eruptions can have complex, multi-component structures broadly analogous to solar CMEs. 
A full stellar CME picture therefore requires simultaneous or nearly simultaneous observations across multiple wavelengths.

\subsection{Radio constraints on CMEs from solar-type stars}\label{sec:10-3}

Radio observations provide a complementary probe of whether eruptive events develop into shocks or escaping plasma at larger heights. 
Promising low-frequency CME-related candidates have recently been reported from M dwarfs, including type-II-like bursts at $\sim 120$--$166$ MHz \citep{2025Natur.647..603C,2025A&A...703A.198K}, whereas comparable radio evidence has not yet been established for solar-type stars. 
For example, the SWAYS campaign found no detectable type-II/III-like burst from EK Draconis in the OVRO-LWA band of 13--87 MHz after a $\sim 4\times10^{34}$ erg superflare \citep{2026arXiv260611492D}. 
Because this frequency range probes plasma emission from low-density material at large heights, the non-detection may indicate that any CME or electron beam did not reach the relevant plasma-frequency layer, possibly at $\gtrsim 10\,R_{\star}$, within the observed time window, did not drive a detectable shock or plasma instability, or produced emission that was too faint or geometrically unfavorable to detect. 
Thus, radio non-detections should be interpreted as constraints on CME propagation, shock formation, and plasma-emission conditions, rather than as simple evidence against eruptions. 
\citet{2026arXiv260611492D} suggest that future radio searches for CMEs from young solar-type stars should combine long-window monitoring, flare-agnostic burst searches, broad frequency coverage, and search strategies guided by stellar coronal and wind models. 
Although this study is not a direct example of multi-wavelength efforts, it provides important guidance for designing future multi-wavelength searches, as discussed in Section~\ref{sec:future}.

\section{Implications for young planetary environments}\label{sec:implication}

Current observations of stellar CME signatures provide empirical constraints for MHD models of CME propagation and plasma-atmosphere interactions, and ultimately for evaluating their cumulative impact on exoplanetary habitability and atmospheric evolution. Around young solar-type stars, planetary atmospheres are affected by both steady-state forcing, such as X-ray/UV radiation and stellar winds, and transient forcing, including flare X-ray/UV radiation, CMEs, and associated energetic particles \citep[e.g.,][]{2012EP&S...64..179L,2020IJAsB..19..136A,2021A&A...649A..96J}. The relative importance of CMEs compared with these other inputs remains poorly constrained because both stellar CME observations and CME propagation/\Add{particle acceleration} models are still incomplete. In this broader context, the possible effects of CMEs on planetary environments can be organized into three broad categories:

\begin{itemize}
    \item \textit{Atmospheric chemistry}: CME-driven shocks and associated energetic particles can ionize and dissociate atmospheric molecules, modifying chemical pathways related to ozone loss, greenhouse-gas production, and prebiotic chemistry.

    \item \textit{Atmospheric escape and erosion}: CME impacts can enhance atmospheric loss by increasing plasma-atmosphere interactions, especially for weakly magnetized or unmagnetized planets.

    \item \textit{Magnetospheric and ionospheric disturbances}: CME sheaths and ejecta can compress planetary magnetospheres, disturb ionospheres, and increase particle precipitation into the upper atmosphere.
\end{itemize}

The early Solar System provides an especially important context for evaluating these coupled effects. 
How the early Earth acquired and maintained surface conditions suitable for the emergence of life remains incompletely understood, and was likely influenced by the evolution of solar input to the planetary environment. 
A related example is the long-standing \textit{faint young Sun paradox}, which highlights the importance of solar evolution \citep{1972Sci...177...52S,2012RvGeo..50.2006F}.
When the Sun was young, the solar magnetic activity, including flares and probably CMEs, is thought to have been much stronger than today. 
Therefore, young-Sun activity may have had both negative and positive effects on early planetary environments. Enhanced XUV radiation and CME impacts can promote atmospheric escape and ozone loss \citep{2010AsBio..10..751S,2019AsBio..19...64T,2020IJAsB..19..136A,2021NatAs...5..298C,2023MNRAS.518.2472R}, while energetic particles can drive chemical reactions that produce greenhouse gases, promote the formation of prebiotic molecules, and enhance surface radiation doses \citep{2016NatGe...9..452A,2019ApJ...881..114Y,2020IJAsB..19..136A,2023Life...13.1103K,2026ApJ..1002L..12K,2025AJ....170...40C,2026IAUS..388..295A}. Thus, stellar activity should not be regarded simply as harmful or beneficial. These effects also depend on planetary atmospheric composition, gravity, magnetic field, orbital distance, and the recurrence frequency of energetic events, highlighting the need for interdisciplinary discussion.

Mars and Venus serve as empirical benchmarks for these star-planet interaction processes, because both are terrestrial planets that presently lack strong intrinsic magnetic fields and therefore respond directly to solar-wind and CME forcing. 
The vulnerability of weakly magnetized terrestrial planets to enhanced stellar-wind and CME forcing, including magnetospheric compression and ionospheric energy deposition, has also been examined in MHD and magnetospheric interaction models for young-Earth conditions and close-in habitable-zone planets around active low-mass stars \citep{2007AsBio...7..167K,2011JGRA..116.1217S,2014ApJ...790...57C}. 
For Mars, MAVEN observations and event-specific models have shown that interplanetary CME and SEP events can produce measurable, transient enhancements in atmospheric escape. 
For example, the March 2015 ICME produced a clear atmospheric and plasma response observed by MAVEN \citep{2015Sci...350.0210J}, while the September 2017 event strongly disturbed the Martian induced magnetosphere and has been modeled to enhance ion escape rates \citep{2018GeoRL..45.8871L,2018GeoRL..45.7248M}.
Such events can compress and restructure the induced magnetosphere and ionosphere and increase ion escape. Relevant non-thermal loss processes include pickup-ion escape, sputtering, and photochemical escape \citep{2016JGRE..121.2364B,2018ApJ...859L..14D}. 
These processes are directly relevant to the denser atmosphere of early Mars, early Venus before its greenhouse divergence, and weakly magnetized rocky exoplanets around magnetically active stars \citep{2021ApJ...916...96A,2025ApJ...994...75S}.

The key question is therefore not whether a single CME can remove a large fraction of an atmosphere, but whether the integrated effects of many CME impacts over geological timescales can contribute significantly to atmospheric evolution \citep{2001Natur.412..237J,2016JGRE..121.2364B,2018A&ARv..26....2L,2019AsBio..19...64T}. 
\Add{Translating stellar CME occurrence rates into planetary exposure also requires accounting for CME trajectories and impact probabilities. Numerical modeling indicates that large-scale stellar magnetic fields can deflect CMEs toward the astrospheric current sheet, making impact probabilities sensitive to the inclination of planetary orbits relative to this sheet \citep{2016ApJ...826..195K}. For the young solar analog $\kappa^{1}$ Ceti, modeling based on the reconstructed magnetic field yielded a CME impact probability of $\sim$30\% for early Venus, Earth, and Mars, about six times an earlier estimate \citep{2019ApJ...886L..37K}.}
Further quantitative planetary-impact models require stellar CME inputs such as occurrence rates, masses, velocities, magnetic fields, and associated energetic-particle spectra and fluences, together with stellar-wind velocities, mass-loss rates, XUV conditions, and their history. The importance of CMEs is likely to vary from system to system. The observational constraints on stellar CME properties reviewed here represent the first generation of empirical inputs for population-level models needed to evaluate cumulative, long-timescale effects across the diversity of exoplanetary systems detected by current and future missions such as TESS, PLATO, and the Habitable Worlds Observatory.


\section{Future prospects}\label{sec:future}

The next step is to move from identifying individual candidates to quantitatively characterizing their physical properties and underlying mechanisms. This is one of the major challenges that stellar CME studies should address over the next decade. Progress will require several complementary directions, because no single wavelength or method can provide a complete picture of stellar CMEs.

\textit{Extension to wider ages and stellar types}: Current observational constraints are still based on a small number of highly active stars at limited evolutionary stages. Future surveys should examine how CME occurrence and properties depend on stellar age, from T Tauri stars to evolved solar-type stars, in order to reconstruct the \textit{Sun in time}. It is also important to extend the comparison across spectral type. Many CME-signature searches have begun with M/K dwarfs \citep[e.g.,][]{1990A&A...238..249H,2016A&A...590A..11V,2024ApJ...961..189N,2025ApJ...979...93K,2021NatAs...5..697V,2022ApJ...936..170L,2025Natur.647..603C,2025A&A...703A.198K,2020ApJ...905...23Z,2024A&A...686A..51M}, whose atmospheric and magnetic conditions differ substantially from those of the Sun. 
Establishing the connection from M dwarfs to solar-type stars is therefore essential for understanding stellar CME physics more broadly and for bridging solar physics and cool-dwarf activity through benchmark solar-type stars.

\textit{XUV and radio observations}: \Add{X-ray and} UV spectroscopy can trace plasma over a wide temperature range ($10^4-10^7$ K) and are therefore well suited for CME searches beyond what is accessible with H$\alpha$ alone \citep{2022ApJ...936..170L,2022ApJ...931...76X,namekata2025natas}. However, this capability has not yet been fully exploited, partly because observational opportunities have so far been limited. 
Future proposed UV missions such as ESCAPE \Add{\citep{2019SPIE11118E..08F,2022JATIS...8a4006F}} and LAPYUTA \citep{2024SPIE13093E..0IT} will be important for filling gaps in EUV wavelength coverage and enabling longer, more flexible monitoring. 
\Add{X-ray Doppler measurements remain scarce, although blueshifts have recently been reported for a giant star and a close binary \citep{2019NatAs...3..742A,2023ApJ...948....9I}. High-resolution X-ray spectroscopy with XRISM \citep{2025JATIS..11d2023I} and the future NewAthena mission \citep{2025NatAs...9...36C} may extend such studies to solar-type stars and reveal further signatures of CMEs.}
Low-frequency radio observations are also essential for testing whether stellar CMEs drive shocks, flux-rope expansions, and star-planet interactions, although promising candidates remain rare \citep{2018ApJ...856...39C,2018ApJ...862..113C,2025Natur.647..603C,2025A&A...703A.198K}. 
Facilities such as LOFAR, GMRT, SKA, and future space- or lunar-based radio observatories will be crucial to search for radio counterparts. 
It is recommended that these facilities prepare observing modes and coordination frameworks for time-domain astronomy across a wide wavelength range.


\textit{Coordinated multi-wavelength campaigns}: Most stellar CME candidates have so far relied on a single wavelength or signature, making their interpretation uncertain. A more robust strategy is to combine complementary diagnostics for the same event, including Doppler-shifted spectral lines, coronal dimming, and radio bursts. The EK Draconis campaigns have already demonstrated the value of this approach by linking H$\alpha$ eruption signatures, possible X-ray dimming, FUV blueshifts, and surface magnetic information \citep{2024ApJ...961...23N,namekata2025natas}.
Although such coordination is challenging, future progress will require stronger community-level links among different wavelength communities, observing facilities, and countries.

\textit{Direct comparison with models}: As more observational constraints have become available, forward radiative hydrodynamics modeling and data-driven MHD simulations are needed to translate observed signatures into physical CME parameters and planetary-impact inputs. Initial efforts have begun for EK Draconis eruptions \citep{2024ApJ...963...50I,2024ApJ...976..255N}, while more sophisticated three-dimensional simulations of stellar eruptions have also advanced rapidly \citep[e.g.,][]{2018ApJ...862...93A,2019ApJ...880...97L,2022SciA....8I9743H,2022ApJ...928..154J,2025arXiv250404144X}. The next step is to apply such models to individual stars using well-constrained magnetic-field configurations and coronal heating parameters and compare them directly with multi-wavelength observations.
Closer collaboration between observers and theorists is needed.

\textit{Filling gaps between the Sun and stars}: Solar observations should continue to be used as a calibration tool, though there are still gaps between solar and stellar observations. 
Closer collaboration between the solar and stellar communities will be important for examining how individual eruptions appear in multi-wavelength Sun-as-a-star observations, what physical quantities can be extracted from them, and how confined events can be distinguished from escaping CMEs. 

\section{Summary and conclusions}
\label{sec:summary_conclusion}

Recent observations have begun to provide empirical constraints on CMEs from young solar-type stars. 
Dedicated H$\alpha$ monitoring has revealed blue-shifted absorption and emission components associated with superflares, interpreted as on-disk filament eruptions and off-limb prominence eruptions, respectively (Sections~\ref{sec:observational_diagnostics} and \ref{sec:discovery_filament_prominence}). 
Sun-as-a-star comparisons, data-driven modeling, large velocities and spatial scales, and inferred masses of $\sim 10^{17}$--$10^{20}$ g all support the interpretation that at least some of these events are associated with escaping CMEs (Sections~\ref{sec:sun_as_a_star}--\ref{sec:velocity_mass_kinetic_energy}). 
Long-term monitoring, magnetic-field constraints, and multi-wavelength observations further suggest that such eruptions may occur frequently, contribute to stellar mass loss, and have multi-temperature structures broadly analogous to those of solar CMEs (Sections~\ref{sec:frequency}--\ref{sec:multi_wavelength}). 
Taken together, these results suggest that young solar-type stars can produce frequent, fast, and massive CME-related eruptions, providing the first empirical basis for evaluating their role in young stellar and planetary environments.

These results have important implications for young planetary environments (Section \ref{sec:implication}). If young solar-type stars frequently produce fast and massive CMEs, their planets are exposed not only to enhanced XUV radiation and stellar winds, but also to transient impacts from magnetized plasma, shocks, and energetic particles. Such events can contribute to atmospheric escape, magnetospheric and ionospheric disturbances, and particle-driven atmospheric chemistry. 

The next decade should move stellar CME studies from candidate identification to quantitative characterization (Section \ref{sec:future}). This will require larger samples across stellar age and spectral type, coordinated X-ray/UV/optical/radio campaigns, direct comparison with forward and MHD models, and further development of the solar–stellar connection. By combining these approaches, future studies will clarify how superflares produce escaping CMEs, how these eruptions propagate through active stellar coronae, and how strongly they shape the space weather environments of young planets.

\backmatter





\bmhead{Acknowledgements}

\Add{The author thanks the referee for their careful review, which has helped improve the quality of the manuscript.}
The author is deeply grateful to all collaborators who contributed to the studies reviewed in this article, including those involved in observations, data analysis, theoretical modeling, multi-wavelength coordination, and the development of instruments and telescopes.
The author is also grateful to Vladimir Airapetian, Meng Jin and Yuta Notsu for their valuable comments.
The author acknowledges support from the Japan Society for the Promotion of Science (JSPS) Overseas Research Fellowships, JSPS KAKENHI Grant Nos. 21J00316, 24H00248, 24K00680, and 25K01041, and the Operation Management Laboratory (OML) of the National Institutes of Natural Sciences (NINS), Japan.
The author acknowledges ideas from the participants in the workshop ``Blazing Paths to Observing Stellar and Exoplanet Particle Environments" organized by the W.M. Keck Institute for Space Studies.
The author would also like to acknowledge the relevant discussions in the International Space Science Institute (ISSI)
Workshop ``Stellar Magnetism and its Impact on (Exo)Planets (\url{https://workshops.issibern.ch/stellar-magnetism/})".



\section*{Declarations}


\begin{itemize}
\item Funding

This work was supported by the JSPS Overseas Research Fellowships and JSPS KAKENHI Grant Numbers 21J00316, 24H00248, 24K00680, and 25K01041 (K.N.).

\item Conflict of interest/Competing interests

The author declares no competing interests.

\item Ethics approval and consent to participate

Not applicable.

\item Consent for publication

Not applicable.

\item Data availability 

Not applicable.

\item Materials availability

Not applicable.

\item Code availability 

Not applicable.

\item Author contribution

K.N. wrote the manuscript.
\end{itemize}

\bibliography{sn-bibliography}

\end{document}